\documentclass[%
 reprint,
 superscriptaddress,
nofootinbib,
 amsmath,amssymb,
 aps,
 prd,
]{revtex4-2}

\usepackage{graphicx}
\usepackage{dcolumn}
\usepackage{bm}
\usepackage{amsmath}
\usepackage{amssymb}
\usepackage{amsfonts}
\usepackage{mathtools}
\usepackage{url}
\usepackage[dvipsnames]{xcolor}
\usepackage[colorlinks]{hyperref}
\usepackage{float}
\usepackage[inline]{enumitem}
\usepackage{subfigure}
\usepackage{appendix}

\definecolor{burgundy}{rgb}{0.6, 0.0, 0.0}
\definecolor{persianblue}{rgb}{0.11, 0.22, 0.73}
\definecolor{darkblue}{rgb}{0.0,0.0,0.4}
\hypersetup{colorlinks, citecolor=darkblue, linkcolor=persianblue, urlcolor=persianblue}

\def\nat{Nature}

\def\mnras{MNRAS}
\def\apj{ApJ}

\def\apjs{ApJS}
\def\aap{A\&A}

\def\physrep{Phys. Rep.}

\def\ssr{Space Science Reviews}

\def\jcap{JCAP}

\graphicspath{{./}{figures/}}

\begin{document}

\title{MHD decomposition explains diffuse $\gamma$-ray emission in Cygnus X}

\author{Ottavio Fornieri}
\email{ottavio.fornieri@gssi.it, ORCID: 0000-0002-6095-9876}
\affiliation{Deutsches Elektronen-Synchrotron DESY, Platanenallee 6, 15738 Zeuthen, Germany}
\affiliation{Gran Sasso Science Institute, Viale Francesco Crispi 7, 67100 L’Aquila, Italy}
\affiliation{INFN-Laboratori Nazionali del Gran Sasso (LNGS), Via G. Acitelli 22, 67100 Assergi (AQ), Italy}

\author{Heshou Zhang}
\email{heshou.zhang@desy.de, ORCID: 0000-0003-2840-6152}
\affiliation{Deutsches Elektronen-Synchrotron DESY, Platanenallee 6, 15738 Zeuthen, Germany}
\affiliation{Institut f\"ur Physik und Astronomie, Universit\"at Potsdam, Haus 28, Karl-Liebknecht-Str. 24/25, 14476 Potsdam, Germany.}
\affiliation{Istituto Nazionale di Astrofisica (INAF) - Osservatorio Astronomico di Brera, Via E. Bianchi 46, 23807 Merate(LC), Italy.}

\begin{abstract}
Cosmic-ray (CR) diffusion is the result of the interaction of such charged particles against magnetic fluctuations. These fluctuations originate from large-scale turbulence cascading towards smaller spatial scales, decomposed into three different modes, as described by \textit{magneto-hydro-dynamics} (MHD) theory. As a consequence, the description of particle diffusion strongly depends on the model describing the injected turbulence. Moreover, the amount of energy assigned to each of the three modes is in general not equally divided, which implies that diffusion properties might be different from one region to another. Here, motivated by the detection of different MHD modes inside the Cygnus-X star-forming region, we study the 3D transport of CRs injected by two prominent sources within a two-zone model that represents the distribution of the modes. Then, by convolving the propagated CR-distribution with the neutral gas, we are able to explain the $\gamma$-ray diffuse emission in the region, observed by the Fermi-LAT and HAWC Collaborations. Such a result represents an important step in the long-standing problem of connecting the CR observables with the micro-physics of particle transport.
\end{abstract}


\maketitle

\section{Introduction}
More than one hundred years after the discovery of cosmic rays (CRs), the debate about their origin is still largely ongoing. According to the standard paradigm, CRs are accelerated at the shock fronts of Supernova Remnants (SNRs) via \textit{diffusive shock acceleration} (DSA)~\citep{1987PhR...154....1B} and then transported across our Galaxy, undergoing all the physical processes that are effectively described by the transport equation~\citep{1990acr..book.....B}. The fact that the bulk of CRs is originated at SNRs was first proposed in \citet{1934PNAS...20..259B} and since then motivated by energy considerations. In particular, comparing the power injected by the sources of CRs in the Galaxy with that of a typical SNR, we obtain a reasonable $\sim 10\%$ efficiency for the conversion of their energy budget into CRs~\citep{2019IJMPD..2830022G}.


A limitation in such a paradigm is represented by the maximum energy $E_{\mathrm{max}}$ that CRs from SNRs can reach. This $E_{\mathrm{max}}$ is regulated by the number of accelerating cycles that particles undergo bouncing back and forth at the shock front before escaping, and is therefore eventually limited by the efficiency of confinement in the region upstream of the shock. The excitation of resonant instabilities~\citep{1969ApJ...156..445K} can help in this sense, but hardly allows to achieve energies of the order $\sim \mathcal{O}(100\, \mathrm{TeV})$~\citep{1983A&A...118..223L,1983A&A...125..249L}. This value, however, needs to face the experimental evidence of an all-particle spectrum that extends up to a few $\mathrm{PeV}$'s. For this reason, massive stellar clusters (MSCs) have been invoked as alternative accelerating sites for galactic CRs.

On the theory side, MSCs have been studied for this purpose since long time ago~\citep{1983SSRv...36..173C}. The engine causing the acceleration process could come from the collection of \mbox{$\sim \mathcal{O}(100)$} stellar winds confined in the compact cluster $(R_{\mathrm{cluster}} \sim 1 - 5 \, \mathrm{pc})$, injecting kinetic energy at a rate \mbox{$L_{w} \sim 10^{34} - 10^{38} \, \mathrm{erg \cdot s^{-1}}$}~\citep{Ackermann:2011lfa, Seo:2018mef}, forcing charged particles to undergo multiple shocks before being released into the \textit{Interstellar Medium} (ISM). Whether this mechanism results in a continuous injection of particles~\citep{Aharonian:2018oau} or a burst, once the shell excavated by the winds is dissipated~\citep{1:2021xpo}, is still matter of debate, and depends on the adopted acceleration mechanism (see \citet{Bykov:2020zqf} for a recent review on the topic). Regardless, considering all the stars contributing to the wind luminosity across the Galaxy, a (possibly) sizeable but not dominant contribution to the CR flux at Earth could come from MSCs~\citep{Seo:2018mef}.

From the experimental point of view, the observations of Very High Energy (VHE) $(100 \, \mathrm{GeV} \leq E_\gamma \leq 100 \, \mathrm{TeV})$ $\gamma$-ray emission in compact star clusters --- such as Westerlund 1~\citep{HESS:2012qpm}, Westerlund 2~\citep{2018A&A...611A..77Y} and the Cygnus-X star-forming region~\citep{Ackermann:2011lfa, Abeysekara:2021yum} --- have been interpreted as signatures of local PeV accelerators. 
Since they are compatible with the decay of neutral pions originated by the scattering of CR-hadrons off the molecular clouds --- this process generates photons with energy $\langle E_{\gamma} \rangle \simeq 0.1 E_{\mathrm{CR}}$ --- these findings have opened the way in the search for PeV accelerators, the so-called \textit{PeVatrons}. Among the above-mentioned observations, the \textit{cocoon} at the center of the Cygnus-X region has recently received much attention in a broader multi-messenger sense. In fact, the LHAASO~\citep{2021Natur.594...33C} Collaboration reported the $7\sigma$-detection of $\sim 530$ photons with $E_{\gamma} \geq 100 \, \mathrm{TeV}$ from twelve regions with overlapping known sources, including the Cygnus \textit{cocoon}, where the highest-energy event $(E_{\gamma}^{\mathrm{max}} \simeq 1.4 \, \mathrm{PeV})$ was originated. Similarly, Tibet-AS$\gamma$~\citep{TibetASgamma:2021tpz} reports 10 $\gamma$-ray events from the Galactic plane with $E_{\gamma} \geq 398 \, \mathrm{TeV}$ of clear hadronic origin, 4 of which --- including the one at the highest energy, again $(E_{\gamma}^{\mathrm{max}} \simeq 957 \, \mathrm{TeV})$ --- are coming from the Cygnus region. These observations represent the first direct evidence that stellar clusters may be acceleration sites for PeV CRs.

The last important piece of information comes from the identification, within Cygnus X, of regions where different \textit{magneto-hydro-dynamic} (MHD) modes dominate the turbulent spectrum~\citep{2020NatAs...4.1001Z}. Indeed, as it is well known~\citep{2004ppa..book.....K}, after turbulence is injected, energy is transferred to smaller spatial scales, and it is decomposed into \textit{Alvén} (incompressible), \textit{fast}- and \textit{slow-magnetosonic} (compressible) modes. The amount of energy transferred to each of these modes depends on the driving force that turbulence experiences~\citep{PhysRevX.10.031021}, with also the possibility of mode mixing, in specific environments~\citep{2016MNRAS.463.1026L}. Based on the calculation in \citet{PhysRevX.10.031021} and on the analysis of the polarized synchrotron light, in \citet{2020NatAs...4.1001Z} it is found that the turbulent energy is differently partitioned among the modes in different locations of the Cygnus-X region. Since Alfvén modes cascade anisotropically in wave number~\citep{GS1995ApJ...438..763G, PhysRevLett.89.281102} and, consequently, are not able to confine particles below $E_{\mathrm{CR}} \sim 10 \, \mathrm{TeV}$~\citep{10.1093/mnras/stab355}, this evidence has significant implications on CR transport, that is therefore inhomogeneous.

In this paper, we consider the Cygnus-X region and study the detailed propagation of particles injected by the OB2 cluster and a nearby SNR ($\gamma$-Cygni) in a two-zone diffusion model, where the values of the diffusion coefficients are regulated by the different MHD modes dominating the transport. The resulting CR distribution will serve to reproduce the $\gamma$-ray morphology observed in the region. The paper is organized as follows. First, we describe the details of the model that we use in the simulations and show the resulting CR distributions. Then, we convolve such distributions with the neutral gas in the molecular clouds, the targets generating the observed $\gamma$-rays. Finally, we discuss the results and derive our conclusions.

\section{Simulation setup}\label{sec:simulation_setup}

\begin{figure*}[ht]
    \centering
    \includegraphics[width=0.95\textwidth,height=0.3\textwidth]{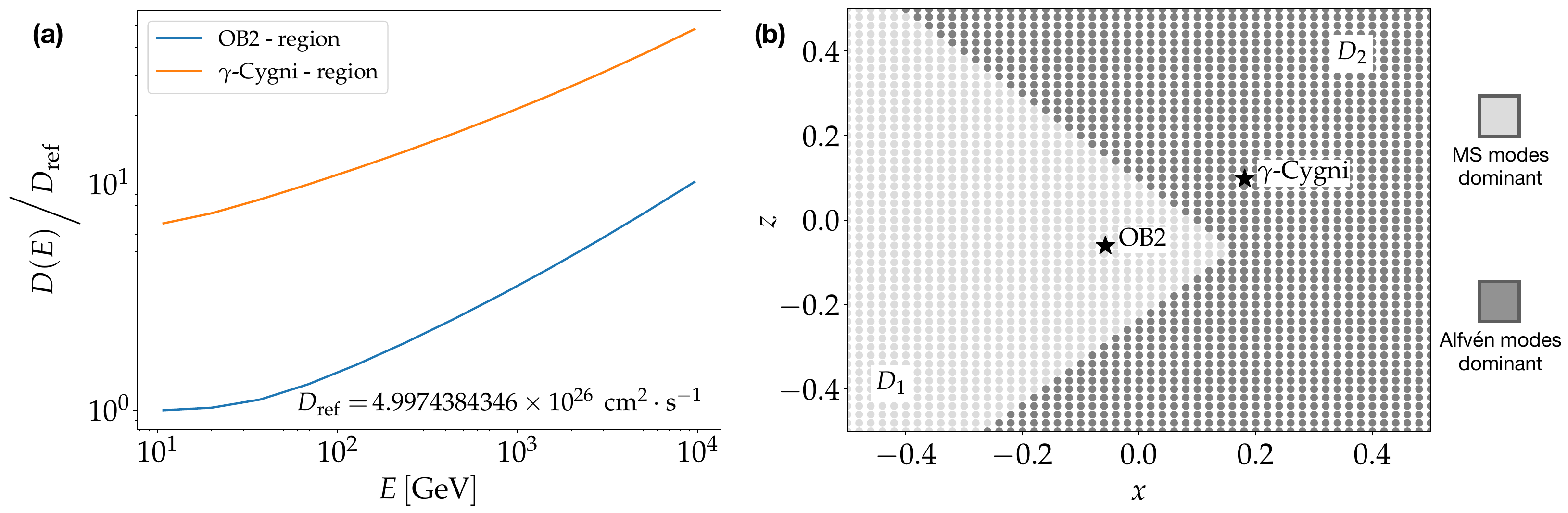}
\caption{\small{The diffusion coefficients in the region are shown. In (a) we report their scaling, according to the parametrization described in the text. In (b) we report their configuration: in particular the $x$ and $z$ axes are, respectively, longitude and latitude of a $(200 \, \mathrm{pc}) \times (200 \, \mathrm{pc})$ region with Galactic coordinates $l^{\mathrm{Sim}}_{\mathrm{Cyg}} \simeq [75.4^\circ, \, 83.6^\circ]$ and $b^{\mathrm{Sim}}_{\mathrm{Cyg}} \simeq [-2.6^\circ, \, 5.6^\circ]$.}}
\label{fig:diffusion_coefficients}
\end{figure*}

\subsection{Sources of CRs in the region} As mentioned in the introduction, much attention has been given to the Cygnus-X region, especially motivated by the possible presence of an accelerator of $\mathrm{PeV}$ CRs. The invoked acceleration mechanism involves the dynamics of stellar winds~\citep{1983SSRv...36..173C, 1:2021xpo} driven by the presence of the OB2 cluster, a young $(t^{\mathrm{OB2}}_{\mathrm{age}} \sim 1-4 \, \mathrm{Myr})$ globular cluster of $\sim \mathcal{O}(100)$ type-O stars dominating the emission in the region --- $\sim 90\%$ of the emission is estimated to come from this association, at $\mathrm{TeV}$ energy as well as in the lower Fermi domain. This region is identified to be HAWC J2030+409 by the HAWC Collaboration~\citep{Abeysekara:2021yum} and is considered to be the counterpart of the $\mathrm{GeV}$ \textit{cocoon} observed by Fermi~\citep{Ackermann:2011lfa}. Another source contributes in the region to the $\gamma$-ray analysis, $\gamma$-Cygni --- 2HWC J2020+403~\citep{Abeysekara:2021yum} likely associated with the VERITAS source VER J2019+407~\citep{2013ApJ...770...93A} ---, a young SNR whose age is estimated from its internal pulsar to be $t^{\mathrm{SNR}}_{\mathrm{age}} \simeq 77 \, \mathrm{kyr}$. The SNR accelerates CRs at the \textit{forward} shock and releases them into the ISM at the beginning of the Sedov-Taylor phase $(t_{\mathrm{Sed}} \leq 10^3 \, \mathrm{yr})$ as a \textit{delta} function in time~\citep{2012JCAP...01..011B}. For what concerns the star cluster, on the other hand, the responsible acceleration mechanism considers a \textit{reverse} shock that traps particles in the inner region for as long as $\sim 1 \, \mathrm{Myr}$, until the shock is dissipated and CRs of all energies are released in the ISM~\citep{1:2021xpo}. In what follows, we consider the physical situation where the OB2 cluster --- with Galactic coordinates $(l_{\mathrm{OB2}}, \, b_{\mathrm{OB2}}) = (80^{\circ}, \, 1^{\circ})$ --- has an age $t_{\mathrm{age}}^{\mathrm{OB2}} = 2 \, \mathrm{Myr}$ and it injected particles $t^{\mathrm{OB2}}_{\mathrm{rel}} = 1.2 \, \mathrm{Myr}$ ago. Additionally, the SNR --- with coordinates $(l_{\mathrm{SNR}}, \, b_{\mathrm{SNR}}) = (78^{\circ}, \, 2.3^{\circ})$ --- injects particles as well after a long time, $t^{\mathrm{SNR}}_{\mathrm{age}} \simeq t^{\mathrm{SNR}}_{\mathrm{rel}} = 77 \, \mathrm{kyr}$ ago, normalized such that $n_{\mathrm{OB2}} \big/ n_{\mathrm{SNR}} = 100$.

\subsection{Transport properties} In order to reproduce the $\gamma$-ray diffuse emission observed by Fermi-LAT~\citep{Ackermann:2011lfa} --- in the range $1 \, \mathrm{GeV} \leq E_{\gamma} \leq 100 \, \mathrm{GeV}$ --- and HAWC~\citep{Abeysekara:2021yum} --- above $E_{\gamma} = 1 \, \mathrm{TeV}$ ---, we propagate parent CRs with energies $10 \, \mathrm{GeV} \leq E_{\mathrm{CR}} \leq 10 \, \mathrm{TeV}$, since we expect the main photon production to be of hadronic origin (see details below), due to neutral pion decay, for which $\langle E_{\gamma} \rangle \simeq 0.1 E_{\mathrm{CR}}$. Due to the declining source spectra of the type $dN_{\mathrm{CR}} / dE \propto E^{-\Gamma_{\mathrm{inj}}}$, with $\Gamma_{\mathrm{inj}} > 2$, the contribution to the final maps coming from more energetic CRs can be considered negligible. For what concerns the nature of the particles responsible for the photon emission, there are clues pointing towards a hadronic origin. Above the $\mathrm{TeV}$ scale this is well-established, as discussed for instance in \citet{Aharonian:2018oau, TibetASgamma:2021tpz}. In the $\mathrm{GeV}$ domain the situation is different: the Radio and X-ray emission constrains the higher-energy $\gamma$-ray data to be not of leptonic origin, as clearly shown in \citet{Abeysekara:2021yum}. It is worth noticing however that, although this implies that a single lepton population cannot be responsible for the \textit{whole} spectrum from Radio to $\gamma$-rays, still it cannot rule out the possibility of an additional leptonic component contributing below $E_{\gamma} \sim 100 \, \mathrm{GeV}$ and then becoming subdominant due to the large magnetic field in the region (see details below) and the consequent rapid loss rates. In what follows, we investigate the hadronic scenario and its implications, leaving the study of a possible lepton contamination to a future work.

In order to propagate CR-protons in the region, we use the findings discussed in \citet{2020NatAs...4.1001Z}, in particular regarding (i) the emerging magnetic field directions and (ii) the different modes dominating different regions in the Cygnus-X area. (i) Regarding the former, there is evidence for a randomly distributed direction of the \textit{total} field $\bm{B}_\mathrm{tot} = \bm{B}_0 + \delta \bm{B}$, being $\bm{B}_0$ and $\delta \bm{B}$ the regular and the turbulent fields, respectively. This implies a 3D particle transport isotropic with respect to $\bm{B}_{\mathrm{tot}}$, so that for the spatial diffusion coefficient it holds $D_{\parallel} \approx D_{\perp} \equiv D$. (ii) Regarding the latter, as mentioned in the introduction, it was found in \citet{2020NatAs...4.1001Z} that magnetosonic modes dominate the area surrounding the Cygnus \textit{cocoon} $(l \simeq [79^\circ, 81^\circ], b \simeq [0^\circ, 2^\circ])$, whereas, on the bottom right $(l \simeq [77^\circ, 80^\circ], b \simeq [-3^\circ, 0^\circ])$ and on the top right $(l \simeq [79^\circ, 81^\circ], b \simeq [1^\circ, 4^\circ])$ of the region, Alfv\'en modes become the main component of turbulence. As a consequence, CR diffusion in the region is isotropic but inhomogeneous. In particular, here we adopt a two-zone diffusion model, where fast modes dominate the region around the \textit{cocoon} (hereafter region 1) and Alfv\'en modes dominate the other areas (hereafter region 2, as they have the same energetics). This partition affects the resulting diffusion coefficients $D_1$ and $D_2$ as described below.

\begin{figure*}
    \centering
    \includegraphics[width=0.9\textwidth,height=0.6\textwidth]{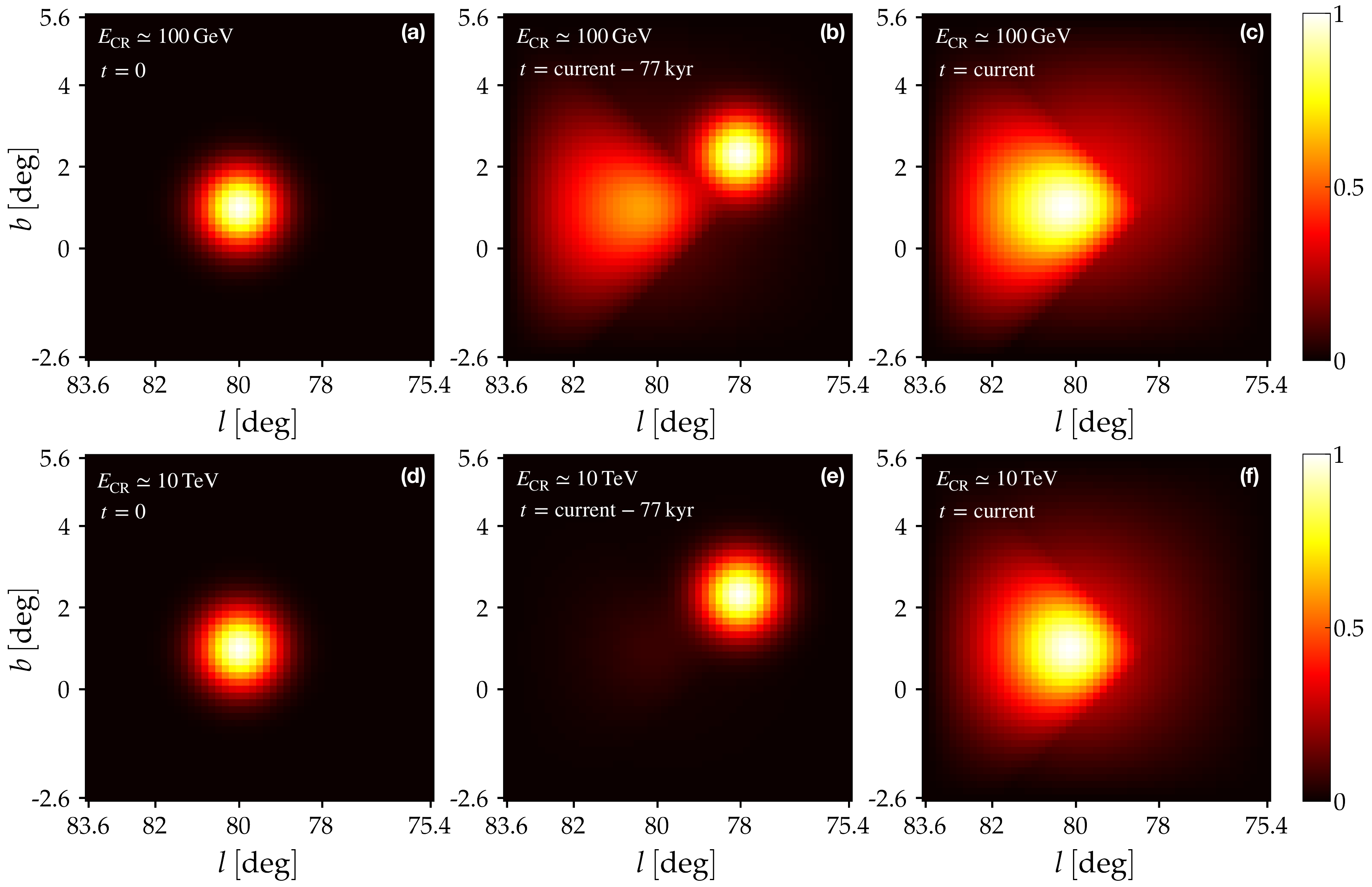}
    \caption{\small{The PoS CR-distribution in the region, caused by the two sources (OB2 and $\gamma$-Cygni) is shown, for different energies (in the rows) and different time-steps of the simulation (in the columns). The particle density is normalized to 1.}}
    \label{fig:CR_distributions_different_energies}
\end{figure*}

It has been shown that, in a developed turbulent cascade, fast modes play a dominant role in confining CRs with respect to Alfv\'en and slow modes~\citep{PhysRevLett.89.281102, Yan:2004aq, Yan:2007uc, 10.1093/mnras/stab355}. In particular, their pitch-angle coefficients is $D^{\mathrm{fast}}_{\mu \mu} \gg D^{\mathrm{A}}_{\mu \mu}$, so that the spatial diffusion coefficient reads 
\begin{equation}\label{eq:spatial_diff_coeff}
\begin{aligned}
    D(E) &= \frac{1}{4} \, \int_{0}^{1} d\mu \, \frac{c^2 \, (1 - \mu^2)^2}{D^{\mathrm{fast}}_{\mu \mu} + D^{\mathrm{slow}}_{\mu \mu} + D^{\mathrm{A}}_{\mu \mu}} \\[6pt]
    &\approx \frac{1}{4} \, \int_{0}^{1} d\mu \, \frac{c^2 \, (1 - \mu^2)^2}{D^{\mathrm{fast}}_{\mu \mu}},
\end{aligned}
\end{equation}
for CRs with pitch-angle cosine $\cos \theta \equiv \mu$, moving at the speed of light $c$. In this context, we remark that the strongly hierarchical relation between $D_{\mu \mu}^{\mathrm{fast}}$ and the other $D_{\mu \mu}$'s leads us to believe that, in our environment, the mixing of the modes does not play a significant role in shaping our diffusion coefficients. The normalized energy density of the turbulent magnetic field, at the injection scale $L$, is $\left(\delta B_L^2 \big/ B_0^2 \right)\big\vert_{\mathrm{tot}} \approx M^2_A$. This quantity can be expressed as the sum of the energy densities in the three modes: $\left( \delta B_L^2 \big/ B_0^2 \right)\big\vert_{\mathrm{tot}} = \left( \delta B_L^2 \big/ B_0^2 \right)\big\vert_{\mathrm{fast}} + \left( \delta B_L^2 \big/ B_0^2 \right)\big\vert_{\mathrm{slow}} + \left( \delta B_L^2 \big/ B_0^2 \right)\big\vert_{\mathrm{Alf}}$. We assume the turbulence in the two regions be trans-Alfv\'enic --- $\left( \delta B_L^2 \big/ B_0^2 \right)\big\vert_{\mathrm{tot}} \sim 1$ --- with a different energy partition among the plasma modes. Within this partition, we are interested in the amount of \textit{fast} modes only, due to the approximation in Equation~\eqref{eq:spatial_diff_coeff}. In particular, in region 1 we assume the fast modes to be the only component, so that $\left( \delta B_L^2 \big/ B_0^2 \right)\big\vert_{\mathrm{fast}} \sim ( M_A^{\mathrm{fast,1}} )^2 \sim 1$, while in region 2, according to \citet{PhysRevX.10.031021}, the energy assigned to the fast modes is reduced to an amount $\lesssim 25\%$, due to the acting driving force. Therefore, the fast-modes magnetic energy in region 2 is $\left( \delta B_L^2 \big/ B_0^2 \right)\big\vert_{\mathrm{fast}} \sim 25\%$, which results in $\left( \delta B_L \big/ B_0 \right)\big\vert_{\mathrm{fast}} \simeq M_A^{\mathrm{fast,2}} = 0.5$. A lower amount would even improve the goodness of our results. 

The diffusion coefficients $D_1$ $(M_A^{\mathrm{fast}} = 1)$ and $D_2$ $(M_A^{\mathrm{fast}} = 0.5)$ are computed according to Equation \eqref{eq:spatial_diff_coeff}, with the code already used for the propagation studies in \citet{10.1093/mnras/stab355}, that can be found at \citet{ottavio_fornieri_2020_4250807}. For the calculation, we use that the local ISM has density $n_{\mathrm{Cyg}} \simeq 10^{-1} \, \mathrm{cm^{-3}}$ and temperature $T_{\mathrm{Cyg}} \simeq 5000 \, \mathrm{K}$~\citep{1978ApJS...36..595D,2005MNRAS.362..424W}, and that the injection scale of the turbulence is reduced with respect to the typical values $\sim \mathcal{O}(100 \, \mathrm{pc})$ invoked for the ISM, due to the compact nature of our sources. Finally, a magnetic-field intensity $B_0 = 20 \, \mu \mathrm{G}$ is considered in \citet{Ackermann:2011lfa, Abeysekara:2021yum}, while a lower one, in the range $B_{0} \sim 5 - 10 \, \mu \mathrm{G}$, is estimated based on the ISM environment in star-forming regions \citep{2005LNP...664..137H, Aharonian:2018oau}. For the sake of definiteness, we consider $B_0 = 10 \, \mu \mathrm{G}$. With our environmental parameters $(n_{\mathrm{Cyg}}, \, T_{\mathrm{Cyg}})$, both \textit{collisional} (viscous) and \textit{collisionless} damping mechanisms for the turbulent waves are taken into account, as discussed in \citet{Yan:2007uc, 10.1093/mnras/stab355}. Motivated by these considerations, for our runs we fix \mbox{$\beta \approx 3.3 \left( 3\, \mu \mathrm{G} \big/ B_{\mathrm{Cyg}} \right)^2 \left( n_{\mathrm{Cyg}} \big/ \mathrm{cm^{-3}} \right) \left( T_{\mathrm{Cyg}} \big/ 10^4 \, \mathrm{K} \right) \sim 0.01$}, and $L_\mathrm{inj} = 10 \, \mathrm{pc}$. The resulting coefficients for the two regions, as a function of the CR energy, are shown in Figure \ref{fig:diffusion_coefficients}a. The ratio between them is $D_2/D_1 \sim 7.5$ at $E_{\mathrm{CR}} \simeq 100 \, \mathrm{GeV}$, and $D_2/D_1 \sim 4.7$ at $E_{\mathrm{CR}} \simeq 10 \, \mathrm{TeV}$.

In the energy range under study, hadronic transport is dominated by diffusion, as we can neglect energy losses, contributing at $E_{\mathrm{CR}} \geq 100 \, \mathrm{TeV}$, and spallation reactions, contributing at $E_{\mathrm{CR}} \leq 1 \, \mathrm{GeV}$. Therefore, we compute the CR transport by solving the following 3D diffusion equation for each energy, from \mbox{$E_{\mathrm{CR}} = 10 \, \mathrm{GeV}$} to \mbox{$E_{\mathrm{CR}} = 10 \, \mathrm{TeV}$}:
\begin{equation}\label{eq:diffusion_equation_3D}
\begin{aligned}
    \frac{\partial n(\bm{r})}{\partial t} = \sum_{i=1,2,3}\frac{\partial}{\partial x_i} \left( D(\bm{r}) \, \frac{\partial n(\bm{r})}{\partial x_i} \right) + \mathcal{S}(\bm{r}, t)
\end{aligned}
\end{equation}
where $\bm{r} = (x_1, x_2, x_3) \equiv (x, y, z)$, with boundary conditions $n = 0$ for all times $t$ at the edge of the simulation box and initial condition $n(x, y, z, t=0) = 0$.

In the above equation, $n(x, y, z, t)$ represents the CR density in arbitrary units, and $D(x, y, z)$ the isotropic, inhomogeneous diffusion coefficient, that can take values $D_1$ or $D_2$. The source term $\mathcal{S}(x, y, z, t)$ parametrizes the two sources described above: for their spatial profile, we consider two Gaussians with \textit{full-width-half-maximum} (FWHM) $\sigma_{\mathrm{FWHM}} = 16 \, \mathrm{pc}$, while the energy part is a single power law scaling as $dN_{\mathrm{CR}} \big/ dE \propto E^{-2.1}$, without any sign of cutoff up to the highest energy propagated, injecting nearly all $(\sim 90\%)$ CRs in a time burst. To solve Equation \eqref{eq:diffusion_equation_3D}, we use a \textit{Forward-Euler} explicit numerical scheme, as described in \citet{langtangen2017finite}.

The region of interest has Galactic coordinates $l_{\mathrm{Cyg}} = [77^\circ, \, 82^\circ]$ and $b_{\mathrm{Cyg}} = [-1^\circ, \, 4^\circ]$. We convert this information to build a 3D cube in Cartesian coordinates, as well as the coordinates of the two sources, via the usual transformation equations that consider the Galactic Center to be at $(x_{\mathrm{GC}}, \, y_{\mathrm{GC}}, \, z_{\mathrm{GC}}) = (0, \, 0, \, 0)$ and the Solar System at $(x_{\odot}, \, y_{\odot}, \, z_{\odot}) = (-8, \, 0, \, 0) \, \mathrm{kpc}$:
\begin{equation}\label{eq:gal_to_cart}
\begin{aligned}
    &x = d_{\mathrm{Cyg}} \cdot \cos b \cos l - R_{\odot}\\
    &y = d_{\mathrm{Cyg}} \cdot \cos b \sin l\\
    &z = d_{\mathrm{Cyg}} \cdot \sin b,
\end{aligned}
\end{equation}
where $R_{\odot} = 8 \, \mathrm{kpc}$ and the average distance of the Cygnus-X region $d_{\mathrm{Cyg}} = 1.4 \, \mathrm{kpc}$~\citep{Abeysekara:2021yum}.

\begin{figure*}
    \centering
    \includegraphics[width=0.95\textwidth,height=0.3\textwidth]{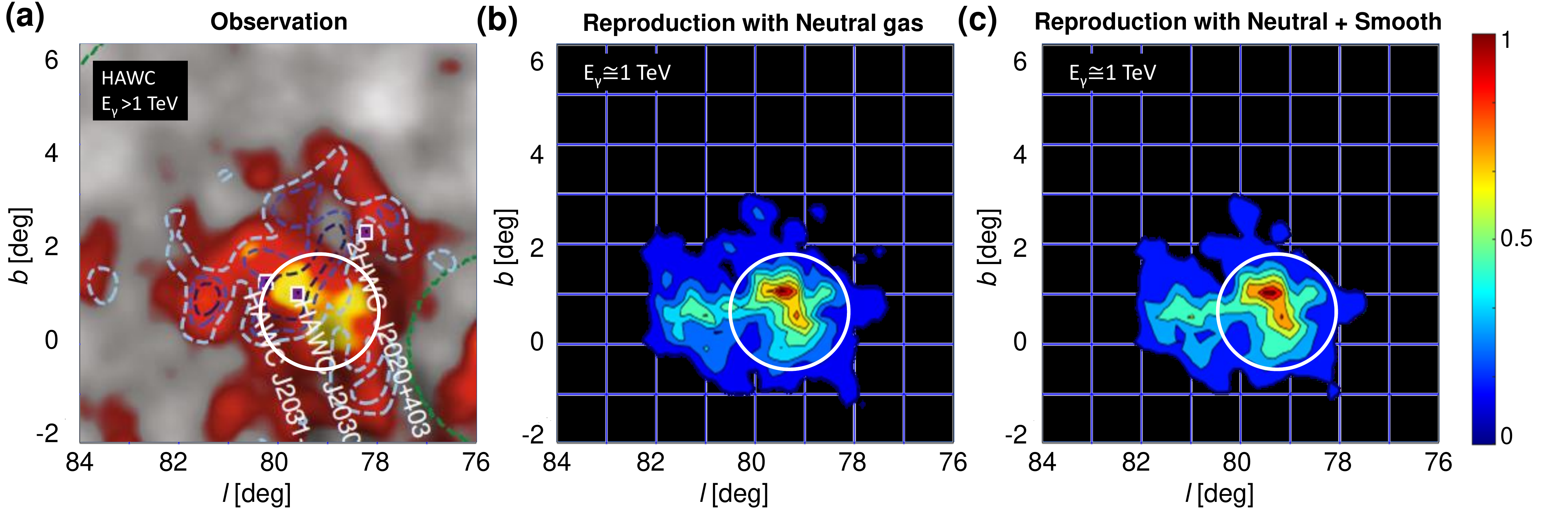}
    \caption{\small{The $\gamma$-ray map at $E_{\gamma} \geq 1 \, \mathrm{TeV}$. The original HAWC observation is shown in (a), where a triangle-shaped enhancement is highlighted as a white circle. The CR interactions with neutral molecules are shown in (b) - (c), with and without the smoothing procedure described in the text. The color bar for (b) and (c) represents the ratio $N_\gamma \big/ N_{\mathrm{max}}$.}}
    \label{fig:gamma_reproduce_HAWC}
\end{figure*}

With the transformation equations \eqref{eq:gal_to_cart}, we map the Plane-of-Sky (PoS) locations of the objects under study into a 3D cuboid, such that $(\Delta x_{\mathrm{Cyg}}, \, \Delta y_{\mathrm{Cyg}}, \, \Delta z_{\mathrm{Cyg}}) \simeq (120,\, 25, \, 122) \, \mathrm{pc}$, $(x_{\mathrm{OB2}}, \, y_{\mathrm{OB2}}, \, z_{\mathrm{OB2}}) \simeq (-7.76, \, 1.38, \, 0.02) \, \mathrm{kpc}$ and $(x_{\mathrm{SNR}}, \, y_{\mathrm{SNR}}, \, z_{\mathrm{SNR}}) \simeq (-7.71, \, 1.37, \, 0.06) \, \mathrm{kpc}$. We can easily notice that, since the $\hat{y}$ coordinates of the region --- and thus of the two sources within it --- is much more distant from the Earth than the $\hat{x}$ and $\hat{z}$ coordinates, the PoS corresponds to the $\widehat{xz}$ plane of the cuboid, with an uncertainty proportional to $\sin \left[(-7.75 + 8) \big/ (1.38 - 0) \right] \sim 0.01$, namely at the level of percent. For practical reasons, we want to construct the simulation on an \textit{equilateral} cuboid. In fact, besides being more convenient to control the resolution of the CR distribution in the region, also the $\hat{y}$ coordinates of the two sources are too close to the edge of the cube, which can have an impact on the final solution of the PDE equation. Therefore, we enlarge the simulation box to $(\Delta x^{\mathrm{Sim}}_{\mathrm{Cyg}}, \, \Delta y^{\mathrm{Sim}}_{\mathrm{Cyg}}, \, \Delta z^{\mathrm{Sim}}_{\mathrm{Cyg}}) \simeq (200, \, 200, \, 200) \, \mathrm{pc}$, corresponding to the Galactic coordinates $l^{\mathrm{Sim}}_{\mathrm{Cyg}} \simeq [75.4^\circ, \, 83.6^\circ]$, $b^{\mathrm{Sim}}_{\mathrm{Cyg}} \simeq [-2.6^\circ, \, 5.6^\circ]$. For our 3D grid, we consider a spatial resolution of $\Delta \phi_{\mathrm{res}} \simeq 0.17^\circ$. Based on what just detailed, the PoS configuration of the region is shown in Figure \ref{fig:diffusion_coefficients}b.

With the setup described above, the CR distributions as observed in the PoS are plotted in Figure \ref{fig:CR_distributions_different_energies}, where $E_{\mathrm{CR}} \simeq 100 \, \mathrm{GeV}$ and $E_{\mathrm{CR}} \simeq 10 \, \mathrm{TeV}$ are shown in the first and second row, respectively, for different time steps of the simulation. In particular, in the first column we report the initial step, corresponding to $t^{\mathrm{OB2}}_{\mathrm{rel}} = 1.2 \, \mathrm{Myr}$ ago --- at this stage only the OB2 cluster has released CRs ---; in the second column we show the moment when the SNR starts to release particles as well, namely $t^{\mathrm{SNR}}_{\mathrm{rel}} = t^{\mathrm{SNR}}_{\mathrm{age}} - t^{\mathrm{SNR}}_{\mathrm{Sed}} \approx 77 \, \mathrm{kyr}$ ago; finally in the third column we show the current age. It is interesting to notice that the last time step corresponds to a steady-state condition for the high-energy particles $(E_{\mathrm{CR}} > 1 \, \mathrm{TeV})$ propagating in region 1, where fast modes dominate diffusion, while lower-energy CR-distributions are still evolving with time. On the other hand, for what concerns region 2, CRs that propagate there but are injected by the OB2 are in steady state, while those injected by the SNR are still evolving, due to their very recent release. Such considerations are easily interpreted in terms of the diffusion coefficient $D_1\big\vert_{10 \, \mathrm{GeV}} \simeq 4.99 \cdot 10^{26} \, \mathrm{cm^2 \cdot s^{-1}}$ that particles experience, according to which steady state is achieved by $10 \, \mathrm{GeV}$ particles after $t_{\mathrm{ss}} \sim 3.5 \cdot 10^6 \, \mathrm{yr}$.

\begin{figure*}
    \centering
    \includegraphics[width=0.95\textwidth,height=0.3\textwidth]{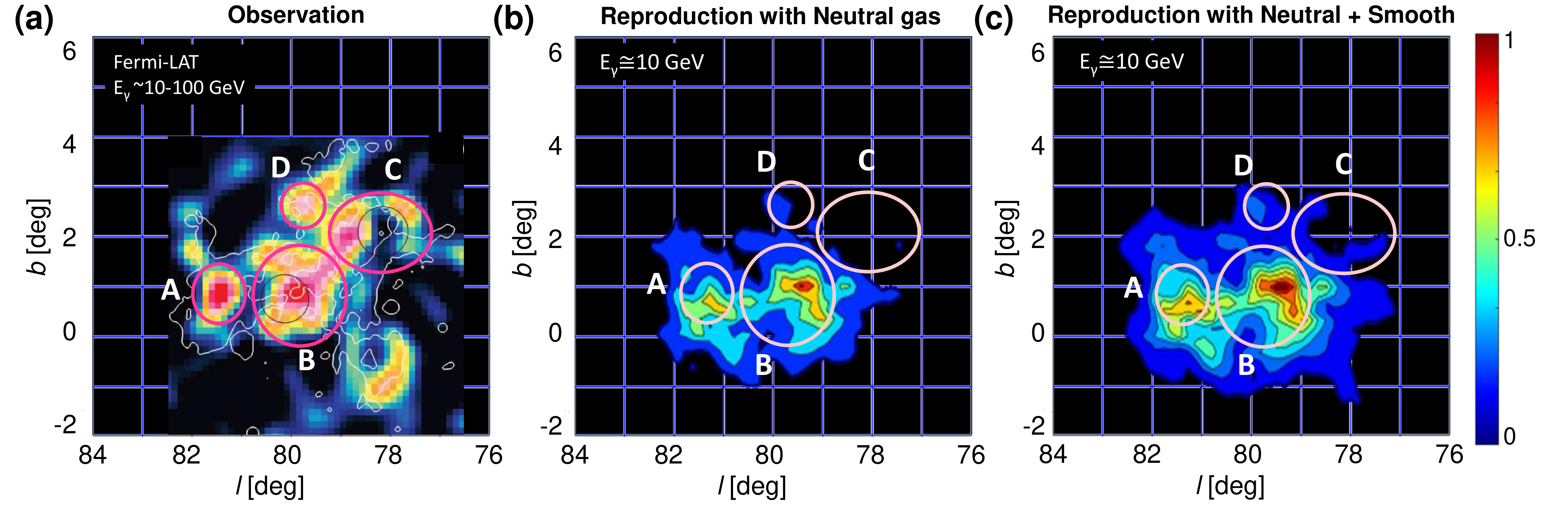}
    \caption{\small{The $\gamma$-ray map at $10 \, \mathrm{GeV} \leq E_{\gamma} \leq 100 \, \mathrm{GeV}$. The original Fermi-LAT observation is shown in (a), where 4 zones are highlighted. The CR interactions with neutral molecules are shown in (b) - (c), with and without the smoothing procedure described in the text. The color bar for (b) and (c) represents the ratio $N_\gamma \big/ N_{\mathrm{max}}$.
    }}
    \label{fig:gamma_reproduce_Fermi}
\end{figure*}

Clearly, particles diffuse in the box according to the properties of the two different regions. In particular, around the \textit{cocoon}, they remain confined for a long time, while they quickly evaporate as they reach the Alfvén-dominated region 2: this is visible in panels (b)-(c)-(f), where an evident separation is present. In panels (b)-(e) --- the instant when $\gamma$-Cygni releases the particles in the box --- the OB2 particles are visible only at low energy (panel (b)), while at high energy (panel (e)) their distribution appears to be very faint compared to the freshly injected SNR-CRs, due to the higher diffusion coefficient the latter experience --- $D_1(E = 10 \, \mathrm{TeV}) \simeq 7.1 \cdot D_1(E = 100 \, \mathrm{GeV})$, as it is shown in Figure \ref{fig:diffusion_coefficients}a. Finally, the last column shows the current configuration of CRs. Since some time has passed from the SNR release, this configuration exclusively reflects the role of particles' diffusivities in the two zones: at $E_{\mathrm{CR}} \simeq 100 \, \mathrm{GeV}$ (panel (c)) CRs propagate slowly, while at $E_{\mathrm{CR}} \simeq 10 \, \mathrm{TeV}$ (panel (f)) they are faster, and the highest-density distribution is concentrated just around the emission area. We verified that an equal injection from the two sources does not notably change the final CR distribution, which therefore is mainly the result of the two-zone diffusion configuration. Due to this last statement, we also note that, even though our CR distribution has been obtained in a purely-Gaussian diffusion regime, contamination due to superdiffusion~\citep{Sampson:2022cju, 10.3389/fspas.2022.900900} are not expected to play a significant role in our result, though the rigidity scalings of the respective diffusion coefficients might be different by a small amount --- especially at low energy $(E_{\mathrm{CR}} \lesssim 1 \, \mathrm{TeV})$.

\section{Results}\label{sec:results}
\subsection{HAWC high-energy $\gamma$-rays} As discussed in the introduction, enhanced emission has been observed by HAWC $(E_{\gamma} \geq 1\, \mathrm{TeV})$ and Fermi-LAT $(E_{\gamma} \simeq 10 - 100 \, \mathrm{GeV})$ and claimed to be of diffusive origin, namely subtracting the contributions from the known sources. In this Section, we investigate such photon production as a result of the hadronic interaction between our simulated cosmic rays and the target gas. We use the CO molecule as a tracer to represent the baryon distribution, which is given as a 2D map~\citep{COcompare}. For each pixel in the map, a 2D-Gaussian distribution is considered, with a $68\%$ containment radius of $\sigma^{\mathrm{CO}}_{2\mathrm{D}} = 0.125^\circ$. Since we have a 2D map for the gas and a 3D CR distribution, we first integrate the latter along the line-of-sight (l.o.s.) and then multiply the result by the baryon density, pixel by pixel. This convolution procedure corresponds to the following calculation:
\begin{equation}\label{eq:raw_convolution}
    N_{\gamma} = n_{\mathrm{CR}} \ast n_{\mathrm{CO}}  = \left( \int_{\mathrm{l.o.s.}} dl \, n_{\mathrm{CR}}(\bm{r}) \right) \cdot n_{\mathrm{CO}} 
\end{equation}
where $N_{\gamma}$ is the number density of the photons and $n_{\mathrm{CO}}$ the column density of the gas.

The resulting \textit{raw} $\gamma$-ray map is given as a pixel-by-pixel distribution. Then, we spread the flux contained in each pixel as to have a 2D Gaussian distribution where the $68\%$ containment radius is $\sigma^\gamma_{2\mathrm{D}} = 0.17^\circ$ (consistent with the HAWC and Fermi measurement), which therefore may overlap with the neighboring pixels. The evolution of the $\gamma$-ray maps for energy \mbox{$E_{\gamma} \geq 1 \, \mathrm{TeV}$} is shown in Figure~\ref{fig:gamma_reproduce_HAWC}. In particular, the HAWC measured emission is shown in panel (a), the \textit{raw} convolution computed with Equation \eqref{eq:raw_convolution} in panel (b), and the final smoothed calculation in panel (c). As is highlighted by the circled region in the figures, we observe a triangle-shaped enhancement at $l_{\mathrm{HAWC}} = [78^\circ, \, 80^\circ], \; b_{\mathrm{HAWC}} = [-0.5^\circ, \, 1.5^\circ]$. Remarkably, even though the gas distribution has an evident enhancement at $(l_{\mathrm{CO}}, b_{\mathrm{CO}}) = (79^\circ, 0^\circ)$~\citep{COcompare}, this region is Alfvén-dominated, which makes CRs escape quickly. On the other hand, the highest intensity in our simulated emission is reached at $(l, b) \simeq (79.5^\circ, 1^\circ)$ --- here fast modes confine particles ---, that very-well resembles the observed structure.

\subsection{Fermi-LAT low-energy $\gamma$-rays} The morphology of the enhanced $\gamma$-ray emission observed by Fermi~\citep{Ackermann:2011lfa} is more extended than that seen by HAWC. In order to guide the comparison, we focus on the 4 areas with the highest significance (pink regions in Figure \ref{fig:gamma_reproduce_Fermi}a): A. the enhancement on the left side $(l_{\mathrm{Fermi,A}} = [80.5^\circ, \, 82^\circ], \; b_{\mathrm{Fermi,A}} = [0.3^\circ, \, 1.5^\circ])$; B. the low-energy counterpart of the HAWC structure, the so-called \textit{cocoon}; C. the extended enhancement in the upper-right near $\gamma$-Cygni $(l_{\mathrm{Fermi,C}} = [77.5^\circ, \, 79.5^\circ], \; b_{\mathrm{Fermi,C}} = [1.5^\circ, \, 3^\circ])$; D. the extended zone on the top $(l_{\mathrm{Fermi,D}} = [79.4^\circ, \, 80.2^\circ], \; b_{\mathrm{Fermi,D}} = [2.2^\circ, \, 3^\circ])$. This configuration is shown in Figure \ref{fig:gamma_reproduce_Fermi}, where in panel (a) the Fermi observation is reported, while in panel (b)-(c) we show the \textit{raw} and smoothed convoluted maps, respectively, as resulting from the procedure described above. As we can see in the figure, the main difference between this morphology and the HAWC map is that, besides the \textit{cocoon} in B, a significant excess is detected in the areas identified as A and D. Considering the more extended area that $E_{\mathrm{CR}} \simeq 100 \, \mathrm{GeV}$ CRs cover with respect to $E_{\mathrm{CR}} \simeq 10 \, \mathrm{TeV}$ ones, such a difference can be explained and these enhancements are reproduced with reasonably good agreement, as seen in Figure~\ref{fig:gamma_reproduce_Fermi}c. On the other hand, for B it similarly holds what we described for the high-intensity gas structure at $(l_{\mathrm{CO}}, b_{\mathrm{CO}}) = (79^\circ, 0^\circ)$. 
The yellow structure at $(l_{\mathrm{CO, 2}}, b_{\mathrm{CO, 2}}) = (78^\circ, -1^\circ)$ --- inside the Alfvén-dominated area --- has, in the CO map, the same significance as the one at $(l_{\mathrm{CO}}, b_{\mathrm{CO}})$ discussed above. This is not matched in the final photon emissivity, hence reflecting the distribution of the plasma modes instead of the gas. The enhanced emission in zone C is not reproduced in the current calculation due to the low CO-intensity here, so that CRs have no target to scatter off. Nonetheless, spots of intense emission in C are present (panel (a)). With this regards, the Fermi map shows an empty circle at the center of zone C, which corresponds to the $\gamma$ rays directly injected by the SNR $\gamma$-Cygni, removed to estimate the diffuse component. As the enhancement around this circle is shaped as a ``ring'', what is considered as diffuse emission in zone C may have likely received some contamination from single-source $\gamma$-rays that were not extracted with the Gaussian cut~\citep{Ackermann:2011lfa}.

In conclusion, the diffuse $\gamma$-ray emission observed by both HAWC and Fermi-LAT is nicely interpreted in terms of hadronic interaction with the gas. Structures are reproduced, which are not correlated with the configuration of the gas, but rather with the distribution of the plasma modes (see Figure 4e in \citet{2020NatAs...4.1001Z}).

\vspace{-0.2cm}
\section{Discussion and Summary}\label{sec:discussion_conclusion}

In this work, we have taken into account the distribution of the plasma modes identified in the Cygnus-X region to build a two-zone diffusion model where we propagated CR-hadrons. We studied the $\gamma$-ray photons produced by the interactions between our simulated CR distribution and the baryon population in the region, which is represented by the CO molecular gas. Our calculations nicely match the $\gamma$-ray observations from HAWC $(\sim 1 \, \mathrm{TeV})$ especially, whereas key structures have been identified in the Fermi-LAT domain $(10 - 100 \, \mathrm{GeV})$. With this respect, we argued that a more refined subtraction procedure for the SNR may be responsible for spurious enhancements in the low-energy band and that a slight contamination from CR leptons cannot be ruled out by current observations. Overall, we find that the hadronic origin of the emission can be considered satisfactory.

Our findings are based on a distribution of plasma modes that locally shape the diffusion coefficient, that is therefore an inhomogeneous property of the medium.

\vspace{-0.2cm}
\section{Acknowledgments}
The authors are grateful to Pasquale Blasi, Pedro De La Torre Luque, Carmelo Evoli, Stefano Gabici, Daniele Gaggero, Tim Linden, Andrew Taylor, Walter Winter, Huirong Yan and the DESY Theory Group for their helpful comments and the deep discussions that led to the realization of this manuscript. H.Z. acknowledges the ERC-granted "Hot Milk" research team.

\vspace{-0.3cm}
\bibliography{apssamp}

\begin{thebibliography}{37}%
\makeatletter
\providecommand \@ifxundefined [1]{%
 \@ifx{#1\undefined}
}%
\providecommand \@ifnum [1]{%
 \ifnum #1\expandafter \@firstoftwo
 \else \expandafter \@secondoftwo
 \fi
}%
\providecommand \@ifx [1]{%
 \ifx #1\expandafter \@firstoftwo
 \else \expandafter \@secondoftwo
 \fi
}%
\providecommand \natexlab [1]{#1}%
\providecommand \enquote  [1]{``#1''}%
\providecommand \bibnamefont  [1]{#1}%
\providecommand \bibfnamefont [1]{#1}%
\providecommand \citenamefont [1]{#1}%
\providecommand \href@noop [0]{\@secondoftwo}%
\providecommand \href [0]{\begingroup \@sanitize@url \@href}%
\providecommand \@href[1]{\@@startlink{#1}\@@href}%
\providecommand \@@href[1]{\endgroup#1\@@endlink}%
\providecommand \@sanitize@url [0]{\catcode `\\12\catcode `\$12\catcode
  `\&12\catcode `\#12\catcode `\^12\catcode `\_12\catcode `\%12\relax}%
\providecommand \@@startlink[1]{}%
\providecommand \@@endlink[0]{}%
\providecommand \url  [0]{\begingroup\@sanitize@url \@url }%
\providecommand \@url [1]{\endgroup\@href {#1}{\urlprefix }}%
\providecommand \urlprefix  [0]{URL }%
\providecommand \Eprint [0]{\href }%
\providecommand \doibase [0]{http://dx.doi.org/}%
\providecommand \selectlanguage [0]{\@gobble}%
\providecommand \bibinfo  [0]{\@secondoftwo}%
\providecommand \bibfield  [0]{\@secondoftwo}%
\providecommand \translation [1]{[#1]}%
\providecommand \BibitemOpen [0]{}%
\providecommand \bibitemStop [0]{}%
\providecommand \bibitemNoStop [0]{.\EOS\space}%
\providecommand \EOS [0]{\spacefactor3000\relax}%
\providecommand \BibitemShut  [1]{\csname bibitem#1\endcsname}%
\let\auto@bib@innerbib\@empty
\bibitem [{\citenamefont {{Blandford}}\ and\ \citenamefont
  {{Eichler}}(1987)}]{1987PhR...154....1B}%
  \BibitemOpen
  \bibfield  {author} {\bibinfo {author} {\bibfnamefont {R.}~\bibnamefont
  {{Blandford}}}\ and\ \bibinfo {author} {\bibfnamefont {D.}~\bibnamefont
  {{Eichler}}},\ }\href {\doibase 10.1016/0370-1573(87)90134-7} {\bibfield
  {journal} {\bibinfo  {journal} {\physrep}\ }\textbf {\bibinfo {volume}
  {154}},\ \bibinfo {pages} {1} (\bibinfo {year} {1987})}\BibitemShut {NoStop}%
\bibitem [{\citenamefont {{Berezinskii}}\ \emph {et~al.}(1990)\citenamefont
  {{Berezinskii}}, \citenamefont {{Bulanov}}, \citenamefont {{Dogiel}},\ and\
  \citenamefont {{Ptuskin}}}]{1990acr..book.....B}%
  \BibitemOpen
  \bibfield  {author} {\bibinfo {author} {\bibfnamefont {V.~S.}\ \bibnamefont
  {{Berezinskii}}}, \bibinfo {author} {\bibfnamefont {S.~V.}\ \bibnamefont
  {{Bulanov}}}, \bibinfo {author} {\bibfnamefont {V.~A.}\ \bibnamefont
  {{Dogiel}}}, \ and\ \bibinfo {author} {\bibfnamefont {V.~S.}\ \bibnamefont
  {{Ptuskin}}},\ }\href@noop {} {\emph {\bibinfo {title} {{Astrophysics of
  cosmic rays}}}}\ (\bibinfo {year} {1990})\BibitemShut {NoStop}%
\bibitem [{\citenamefont {{Baade}}\ and\ \citenamefont
  {{Zwicky}}(1934)}]{1934PNAS...20..259B}%
  \BibitemOpen
  \bibfield  {author} {\bibinfo {author} {\bibfnamefont {W.}~\bibnamefont
  {{Baade}}}\ and\ \bibinfo {author} {\bibfnamefont {F.}~\bibnamefont
  {{Zwicky}}},\ }\href {\doibase 10.1073/pnas.20.5.259} {\bibfield  {journal}
  {\bibinfo  {journal} {Proceedings of the National Academy of Science}\
  }\textbf {\bibinfo {volume} {20}},\ \bibinfo {pages} {259} (\bibinfo {year}
  {1934})}\BibitemShut {NoStop}%
\bibitem [{\citenamefont {Gabici}\ \emph {et~al.}(2019)\citenamefont {Gabici},
  \citenamefont {Evoli}, \citenamefont {Gaggero}, \citenamefont {Lipari},
  \citenamefont {Mertsch}, \citenamefont {Orlando}, \citenamefont {Strong},\
  and\ \citenamefont {Vittino}}]{2019IJMPD..2830022G}%
  \BibitemOpen
  \bibfield  {author} {\bibinfo {author} {\bibfnamefont {S.}~\bibnamefont
  {Gabici}}, \bibinfo {author} {\bibfnamefont {C.}~\bibnamefont {Evoli}},
  \bibinfo {author} {\bibfnamefont {D.}~\bibnamefont {Gaggero}}, \bibinfo
  {author} {\bibfnamefont {P.}~\bibnamefont {Lipari}}, \bibinfo {author}
  {\bibfnamefont {P.}~\bibnamefont {Mertsch}}, \bibinfo {author} {\bibfnamefont
  {E.}~\bibnamefont {Orlando}}, \bibinfo {author} {\bibfnamefont
  {A.}~\bibnamefont {Strong}}, \ and\ \bibinfo {author} {\bibfnamefont
  {A.}~\bibnamefont {Vittino}},\ }\href {\doibase 10.1142/S0218271819300222}
  {\bibfield  {journal} {\bibinfo  {journal} {Int. J. Mod. Phys. D}\ }\textbf
  {\bibinfo {volume} {28}},\ \bibinfo {pages} {1930022} (\bibinfo {year}
  {2019})},\ \Eprint {http://arxiv.org/abs/1903.11584} {arXiv:1903.11584
  [astro-ph.HE]} \BibitemShut {NoStop}%
\bibitem [{\citenamefont {{Kulsrud}}\ and\ \citenamefont
  {{Pearce}}(1969)}]{1969ApJ...156..445K}%
  \BibitemOpen
  \bibfield  {author} {\bibinfo {author} {\bibfnamefont {R.}~\bibnamefont
  {{Kulsrud}}}\ and\ \bibinfo {author} {\bibfnamefont {W.~P.}\ \bibnamefont
  {{Pearce}}},\ }\href {\doibase 10.1086/149981} {\bibfield  {journal}
  {\bibinfo  {journal} {\apj}\ }\textbf {\bibinfo {volume} {156}},\ \bibinfo
  {pages} {445} (\bibinfo {year} {1969})}\BibitemShut {NoStop}%
\bibitem [{\citenamefont {{Lagage}}\ and\ \citenamefont
  {{Cesarsky}}(1983{\natexlab{a}})}]{1983A&A...118..223L}%
  \BibitemOpen
  \bibfield  {author} {\bibinfo {author} {\bibfnamefont {P.~O.}\ \bibnamefont
  {{Lagage}}}\ and\ \bibinfo {author} {\bibfnamefont {C.~J.}\ \bibnamefont
  {{Cesarsky}}},\ }\href@noop {} {\bibfield  {journal} {\bibinfo  {journal}
  {\aap}\ }\textbf {\bibinfo {volume} {118}},\ \bibinfo {pages} {223} (\bibinfo
  {year} {1983}{\natexlab{a}})}\BibitemShut {NoStop}%
\bibitem [{\citenamefont {{Lagage}}\ and\ \citenamefont
  {{Cesarsky}}(1983{\natexlab{b}})}]{1983A&A...125..249L}%
  \BibitemOpen
  \bibfield  {author} {\bibinfo {author} {\bibfnamefont {P.~O.}\ \bibnamefont
  {{Lagage}}}\ and\ \bibinfo {author} {\bibfnamefont {C.~J.}\ \bibnamefont
  {{Cesarsky}}},\ }\href@noop {} {\bibfield  {journal} {\bibinfo  {journal}
  {\aap}\ }\textbf {\bibinfo {volume} {125}},\ \bibinfo {pages} {249} (\bibinfo
  {year} {1983}{\natexlab{b}})}\BibitemShut {NoStop}%
\bibitem [{\citenamefont {{Cesarsky}}\ and\ \citenamefont
  {{Montmerle}}(1983)}]{1983SSRv...36..173C}%
  \BibitemOpen
  \bibfield  {author} {\bibinfo {author} {\bibfnamefont {C.~J.}\ \bibnamefont
  {{Cesarsky}}}\ and\ \bibinfo {author} {\bibfnamefont {T.}~\bibnamefont
  {{Montmerle}}},\ }\href {\doibase 10.1007/BF00167503} {\bibfield  {journal}
  {\bibinfo  {journal} {\ssr}\ }\textbf {\bibinfo {volume} {36}},\ \bibinfo
  {pages} {173} (\bibinfo {year} {1983})}\BibitemShut {NoStop}%
\bibitem [{\citenamefont {Ackermann}\ \emph {et~al.}(2011)\citenamefont
  {Ackermann} \emph {et~al.}}]{Ackermann:2011lfa}%
  \BibitemOpen
  \bibfield  {author} {\bibinfo {author} {\bibfnamefont {M.}~\bibnamefont
  {Ackermann}} \emph {et~al.},\ }\href {\doibase 10.1126/science.1210311}
  {\bibfield  {journal} {\bibinfo  {journal} {Science}\ }\textbf {\bibinfo
  {volume} {334}},\ \bibinfo {pages} {1103} (\bibinfo {year}
  {2011})}\BibitemShut {NoStop}%
\bibitem [{\citenamefont {Seo}\ \emph {et~al.}(2018)\citenamefont {Seo},
  \citenamefont {Kang},\ and\ \citenamefont {Ryu}}]{Seo:2018mef}%
  \BibitemOpen
  \bibfield  {author} {\bibinfo {author} {\bibfnamefont {J.}~\bibnamefont
  {Seo}}, \bibinfo {author} {\bibfnamefont {H.}~\bibnamefont {Kang}}, \ and\
  \bibinfo {author} {\bibfnamefont {D.}~\bibnamefont {Ryu}},\ }\href {\doibase
  10.5303/JKAS.2018.51.2.37} {\bibfield  {journal} {\bibinfo  {journal} {J.
  Korean Astron. Soc.}\ }\textbf {\bibinfo {volume} {51}},\ \bibinfo {pages}
  {37} (\bibinfo {year} {2018})},\ \Eprint {http://arxiv.org/abs/1804.07486}
  {arXiv:1804.07486 [astro-ph.HE]} \BibitemShut {NoStop}%
\bibitem [{\citenamefont {Aharonian}\ \emph {et~al.}(2019)\citenamefont
  {Aharonian}, \citenamefont {Yang},\ and\ \citenamefont {de~O\~na
  Wilhelmi}}]{Aharonian:2018oau}%
  \BibitemOpen
  \bibfield  {author} {\bibinfo {author} {\bibfnamefont {F.}~\bibnamefont
  {Aharonian}}, \bibinfo {author} {\bibfnamefont {R.}~\bibnamefont {Yang}}, \
  and\ \bibinfo {author} {\bibfnamefont {E.}~\bibnamefont {de~O\~na
  Wilhelmi}},\ }\href {\doibase 10.1038/s41550-019-0724-0} {\bibfield
  {journal} {\bibinfo  {journal} {Nature Astron.}\ }\textbf {\bibinfo {volume}
  {3}},\ \bibinfo {pages} {561} (\bibinfo {year} {2019})},\ \Eprint
  {http://arxiv.org/abs/1804.02331} {arXiv:1804.02331 [astro-ph.HE]}
  \BibitemShut {NoStop}%
\bibitem [{\citenamefont {Morlino}\ \emph {et~al.}(2021)\citenamefont
  {Morlino}, \citenamefont {Blasi}, \citenamefont {Peretti},\ and\
  \citenamefont {Cristofari}}]{1:2021xpo}%
  \BibitemOpen
  \bibfield  {author} {\bibinfo {author} {\bibfnamefont {G.}~\bibnamefont
  {Morlino}}, \bibinfo {author} {\bibfnamefont {P.}~\bibnamefont {Blasi}},
  \bibinfo {author} {\bibfnamefont {E.}~\bibnamefont {Peretti}}, \ and\
  \bibinfo {author} {\bibfnamefont {P.}~\bibnamefont {Cristofari}},\ }\href
  {\doibase 10.1093/mnras/stab690} {\  (\bibinfo {year} {2021}),\
  10.1093/mnras/stab690},\ \Eprint {http://arxiv.org/abs/2102.09217}
  {arXiv:2102.09217 [astro-ph.HE]} \BibitemShut {NoStop}%
\bibitem [{\citenamefont {Bykov}\ \emph {et~al.}(2020)\citenamefont {Bykov},
  \citenamefont {Marcowith}, \citenamefont {Amato}, \citenamefont {Kalyashova},
  \citenamefont {Kruijssen},\ and\ \citenamefont {Waxman}}]{Bykov:2020zqf}%
  \BibitemOpen
  \bibfield  {author} {\bibinfo {author} {\bibfnamefont {A.~M.}\ \bibnamefont
  {Bykov}}, \bibinfo {author} {\bibfnamefont {A.}~\bibnamefont {Marcowith}},
  \bibinfo {author} {\bibfnamefont {E.}~\bibnamefont {Amato}}, \bibinfo
  {author} {\bibfnamefont {M.~E.}\ \bibnamefont {Kalyashova}}, \bibinfo
  {author} {\bibfnamefont {J.~M.~D.}\ \bibnamefont {Kruijssen}}, \ and\
  \bibinfo {author} {\bibfnamefont {E.}~\bibnamefont {Waxman}},\ }\href
  {\doibase 10.1007/s11214-020-00663-0} {\bibfield  {journal} {\bibinfo
  {journal} {Space Sci. Rev.}\ }\textbf {\bibinfo {volume} {216}},\ \bibinfo
  {pages} {42} (\bibinfo {year} {2020})},\ \Eprint
  {http://arxiv.org/abs/2003.11534} {arXiv:2003.11534 [astro-ph.HE]}
  \BibitemShut {NoStop}%
\bibitem [{\citenamefont {Abramowski}\ \emph {et~al.}(2012)\citenamefont
  {Abramowski} \emph {et~al.}}]{HESS:2012qpm}%
  \BibitemOpen
  \bibfield  {author} {\bibinfo {author} {\bibfnamefont {A.}~\bibnamefont
  {Abramowski}} \emph {et~al.} (\bibinfo {collaboration} {H.E.S.S.}),\ }\href
  {\doibase 10.1051/0004-6361/201117928} {\bibfield  {journal} {\bibinfo
  {journal} {Astron. Astrophys.}\ }\textbf {\bibinfo {volume} {537}},\ \bibinfo
  {pages} {A114} (\bibinfo {year} {2012})},\ \Eprint
  {http://arxiv.org/abs/1111.2043} {arXiv:1111.2043 [astro-ph.HE]} \BibitemShut
  {NoStop}%
\bibitem [{\citenamefont {{Yang}}\ \emph {et~al.}(2018)\citenamefont {{Yang}},
  \citenamefont {{de O{\~n}a Wilhelmi}},\ and\ \citenamefont
  {{Aharonian}}}]{2018A&A...611A..77Y}%
  \BibitemOpen
  \bibfield  {author} {\bibinfo {author} {\bibfnamefont {R.-z.}\ \bibnamefont
  {{Yang}}}, \bibinfo {author} {\bibfnamefont {E.}~\bibnamefont {{de O{\~n}a
  Wilhelmi}}}, \ and\ \bibinfo {author} {\bibfnamefont {F.}~\bibnamefont
  {{Aharonian}}},\ }\href {\doibase 10.1051/0004-6361/201732045} {\bibfield
  {journal} {\bibinfo  {journal} {\aap}\ }\textbf {\bibinfo {volume} {611}},\
  \bibinfo {eid} {A77} (\bibinfo {year} {2018})},\ \Eprint
  {http://arxiv.org/abs/1710.02803} {arXiv:1710.02803 [astro-ph.HE]}
  \BibitemShut {NoStop}%
\bibitem [{\citenamefont {Abeysekara}\ \emph {et~al.}(2021)\citenamefont
  {Abeysekara} \emph {et~al.}}]{Abeysekara:2021yum}%
  \BibitemOpen
  \bibfield  {author} {\bibinfo {author} {\bibfnamefont {A.~U.}\ \bibnamefont
  {Abeysekara}} \emph {et~al.},\ }\href {\doibase 10.1038/s41550-021-01318-y}
  {\bibfield  {journal} {\bibinfo  {journal} {Nature Astron.}\ }\textbf
  {\bibinfo {volume} {5}},\ \bibinfo {pages} {465} (\bibinfo {year} {2021})},\
  \Eprint {http://arxiv.org/abs/2103.06820} {arXiv:2103.06820 [astro-ph.HE]}
  \BibitemShut {NoStop}%
\bibitem [{\citenamefont {{Cao}}\ \emph {et~al.}(2021)\citenamefont {{Cao}}
  \emph {et~al.}}]{2021Natur.594...33C}%
  \BibitemOpen
  \bibfield  {author} {\bibinfo {author} {\bibfnamefont {Z.}~\bibnamefont
  {{Cao}}} \emph {et~al.},\ }\href {\doibase 10.1038/s41586-021-03498-z}
  {\bibfield  {journal} {\bibinfo  {journal} {\nat}\ }\textbf {\bibinfo
  {volume} {594}},\ \bibinfo {pages} {33} (\bibinfo {year} {2021})}\BibitemShut
  {NoStop}%
\bibitem [{\citenamefont {Amenomori}\ \emph {et~al.}(2021)\citenamefont
  {Amenomori} \emph {et~al.}}]{TibetASgamma:2021tpz}%
  \BibitemOpen
  \bibfield  {author} {\bibinfo {author} {\bibfnamefont {M.}~\bibnamefont
  {Amenomori}} \emph {et~al.} (\bibinfo {collaboration} {Tibet ASgamma}),\
  }\href {\doibase 10.1103/PhysRevLett.126.141101} {\bibfield  {journal}
  {\bibinfo  {journal} {Phys. Rev. Lett.}\ }\textbf {\bibinfo {volume} {126}},\
  \bibinfo {pages} {141101} (\bibinfo {year} {2021})},\ \Eprint
  {http://arxiv.org/abs/2104.05181} {arXiv:2104.05181 [astro-ph.HE]}
  \BibitemShut {NoStop}%
\bibitem [{\citenamefont {{Zhang}}\ \emph {et~al.}(2020)\citenamefont
  {{Zhang}}, \citenamefont {{Chepurnov}}, \citenamefont {{Yan}}, \citenamefont
  {{Makwana}}, \citenamefont {{Santos-Lima}},\ and\ \citenamefont
  {{Appleby}}}]{2020NatAs...4.1001Z}%
  \BibitemOpen
  \bibfield  {author} {\bibinfo {author} {\bibfnamefont {H.}~\bibnamefont
  {{Zhang}}}, \bibinfo {author} {\bibfnamefont {A.}~\bibnamefont
  {{Chepurnov}}}, \bibinfo {author} {\bibfnamefont {H.}~\bibnamefont {{Yan}}},
  \bibinfo {author} {\bibfnamefont {K.}~\bibnamefont {{Makwana}}}, \bibinfo
  {author} {\bibfnamefont {R.}~\bibnamefont {{Santos-Lima}}}, \ and\ \bibinfo
  {author} {\bibfnamefont {S.}~\bibnamefont {{Appleby}}},\ }\href {\doibase
  10.1038/s41550-020-1093-4} {\bibfield  {journal} {\bibinfo  {journal} {Nature
  Astronomy}\ }\textbf {\bibinfo {volume} {4}},\ \bibinfo {pages} {1001}
  (\bibinfo {year} {2020})},\ \Eprint {http://arxiv.org/abs/1808.01913}
  {arXiv:1808.01913 [physics.plasm-ph]} \BibitemShut {NoStop}%
\bibitem [{\citenamefont {{Kulsrud}}(2004)}]{2004ppa..book.....K}%
  \BibitemOpen
  \bibfield  {author} {\bibinfo {author} {\bibfnamefont {R.~M.}\ \bibnamefont
  {{Kulsrud}}},\ }\href@noop {} {\emph {\bibinfo {title} {{Plasma Physics for
  Astrophysics}}}}\ (\bibinfo {year} {2004})\BibitemShut {NoStop}%
\bibitem [{\citenamefont {Makwana}\ and\ \citenamefont
  {Yan}(2020)}]{PhysRevX.10.031021}%
  \BibitemOpen
  \bibfield  {author} {\bibinfo {author} {\bibfnamefont {K.~D.}\ \bibnamefont
  {Makwana}}\ and\ \bibinfo {author} {\bibfnamefont {H.}~\bibnamefont {Yan}},\
  }\href {\doibase 10.1103/PhysRevX.10.031021} {\bibfield  {journal} {\bibinfo
  {journal} {Phys. Rev. X}\ }\textbf {\bibinfo {volume} {10}},\ \bibinfo
  {pages} {031021} (\bibinfo {year} {2020})}\BibitemShut {NoStop}%
\bibitem [{\citenamefont {{Lehmann}}\ \emph {et~al.}(2016)\citenamefont
  {{Lehmann}}, \citenamefont {{Federrath}},\ and\ \citenamefont
  {{Wardle}}}]{2016MNRAS.463.1026L}%
  \BibitemOpen
  \bibfield  {author} {\bibinfo {author} {\bibfnamefont {A.}~\bibnamefont
  {{Lehmann}}}, \bibinfo {author} {\bibfnamefont {C.}~\bibnamefont
  {{Federrath}}}, \ and\ \bibinfo {author} {\bibfnamefont {M.}~\bibnamefont
  {{Wardle}}},\ }\href {\doibase 10.1093/mnras/stw2015} {\bibfield  {journal}
  {\bibinfo  {journal} {\mnras}\ }\textbf {\bibinfo {volume} {463}},\ \bibinfo
  {pages} {1026} (\bibinfo {year} {2016})},\ \Eprint
  {http://arxiv.org/abs/1608.02050} {arXiv:1608.02050 [astro-ph.GA]}
  \BibitemShut {NoStop}%
\bibitem [{\citenamefont {{Goldreich}}\ and\ \citenamefont
  {{Sridhar}}(1995)}]{GS1995ApJ...438..763G}%
  \BibitemOpen
  \bibfield  {author} {\bibinfo {author} {\bibfnamefont {P.}~\bibnamefont
  {{Goldreich}}}\ and\ \bibinfo {author} {\bibfnamefont {S.}~\bibnamefont
  {{Sridhar}}},\ }\href {\doibase 10.1086/175121} {\bibfield  {journal}
  {\bibinfo  {journal} {\apj}\ }\textbf {\bibinfo {volume} {438}},\ \bibinfo
  {pages} {763} (\bibinfo {year} {1995})}\BibitemShut {NoStop}%
\bibitem [{\citenamefont {Yan}\ and\ \citenamefont
  {Lazarian}(2002)}]{PhysRevLett.89.281102}%
  \BibitemOpen
  \bibfield  {author} {\bibinfo {author} {\bibfnamefont {H.}~\bibnamefont
  {Yan}}\ and\ \bibinfo {author} {\bibfnamefont {A.}~\bibnamefont {Lazarian}},\
  }\href {\doibase 10.1103/PhysRevLett.89.281102} {\bibfield  {journal}
  {\bibinfo  {journal} {Phys. Rev. Lett.}\ }\textbf {\bibinfo {volume} {89}},\
  \bibinfo {pages} {281102} (\bibinfo {year} {2002})}\BibitemShut {NoStop}%
\bibitem [{\citenamefont {Fornieri}\ \emph {et~al.}(2021)\citenamefont
  {Fornieri}, \citenamefont {Gaggero}, \citenamefont {Cerri}, \citenamefont
  {De~La Torre~Luque},\ and\ \citenamefont {Gabici}}]{10.1093/mnras/stab355}%
  \BibitemOpen
  \bibfield  {author} {\bibinfo {author} {\bibfnamefont {O.}~\bibnamefont
  {Fornieri}}, \bibinfo {author} {\bibfnamefont {D.}~\bibnamefont {Gaggero}},
  \bibinfo {author} {\bibfnamefont {S.~S.}\ \bibnamefont {Cerri}}, \bibinfo
  {author} {\bibfnamefont {P.}~\bibnamefont {De~La Torre~Luque}}, \ and\
  \bibinfo {author} {\bibfnamefont {S.}~\bibnamefont {Gabici}},\ }\href
  {\doibase 10.1093/mnras/stab355} {\bibfield  {journal} {\bibinfo  {journal}
  {Monthly Notices of the Royal Astronomical Society}\ }\textbf {\bibinfo
  {volume} {502}},\ \bibinfo {pages} {5821} (\bibinfo {year}
  {2021})}\BibitemShut {NoStop}%
\bibitem [{\citenamefont {{Aliu}}\ \emph {et~al.}(2013)\citenamefont {{Aliu}}
  \emph {et~al.}}]{2013ApJ...770...93A}%
  \BibitemOpen
  \bibfield  {author} {\bibinfo {author} {\bibfnamefont {E.}~\bibnamefont
  {{Aliu}}} \emph {et~al.},\ }\href {\doibase 10.1088/0004-637X/770/2/93}
  {\bibfield  {journal} {\bibinfo  {journal} {\apj}\ }\textbf {\bibinfo
  {volume} {770}},\ \bibinfo {eid} {93} (\bibinfo {year} {2013})},\ \Eprint
  {http://arxiv.org/abs/1305.6508} {arXiv:1305.6508 [astro-ph.HE]} \BibitemShut
  {NoStop}%
\bibitem [{\citenamefont {{Blasi}}\ and\ \citenamefont
  {{Amato}}(2012)}]{2012JCAP...01..011B}%
  \BibitemOpen
  \bibfield  {author} {\bibinfo {author} {\bibfnamefont {P.}~\bibnamefont
  {{Blasi}}}\ and\ \bibinfo {author} {\bibfnamefont {E.}~\bibnamefont
  {{Amato}}},\ }\href {\doibase 10.1088/1475-7516/2012/01/011} {\bibfield
  {journal} {\bibinfo  {journal} {\jcap}\ }\textbf {\bibinfo {volume} {2012}},\
  \bibinfo {eid} {011} (\bibinfo {year} {2012})},\ \Eprint
  {http://arxiv.org/abs/1105.4529} {arXiv:1105.4529 [astro-ph.HE]} \BibitemShut
  {NoStop}%
\bibitem [{\citenamefont {Yan}\ and\ \citenamefont
  {Lazarian}(2004)}]{Yan:2004aq}%
  \BibitemOpen
  \bibfield  {author} {\bibinfo {author} {\bibfnamefont {H.}~\bibnamefont
  {Yan}}\ and\ \bibinfo {author} {\bibfnamefont {A.}~\bibnamefont {Lazarian}},\
  }\href {\doibase 10.1086/423733} {\bibfield  {journal} {\bibinfo  {journal}
  {Astrophys. J.}\ }\textbf {\bibinfo {volume} {614}},\ \bibinfo {pages} {757}
  (\bibinfo {year} {2004})},\ \Eprint {http://arxiv.org/abs/astro-ph/0408172}
  {arXiv:astro-ph/0408172} \BibitemShut {NoStop}%
\bibitem [{\citenamefont {Yan}\ and\ \citenamefont
  {Lazarian}(2008)}]{Yan:2007uc}%
  \BibitemOpen
  \bibfield  {author} {\bibinfo {author} {\bibfnamefont {H.}~\bibnamefont
  {Yan}}\ and\ \bibinfo {author} {\bibfnamefont {A.}~\bibnamefont {Lazarian}},\
  }\href {\doibase 10.1086/524771} {\bibfield  {journal} {\bibinfo  {journal}
  {Astrophys. J.}\ }\textbf {\bibinfo {volume} {673}},\ \bibinfo {pages} {942}
  (\bibinfo {year} {2008})},\ \Eprint {http://arxiv.org/abs/0710.2617}
  {arXiv:0710.2617 [astro-ph]} \BibitemShut {NoStop}%
\bibitem [{\citenamefont {Fornieri}(2020)}]{ottavio_fornieri_2020_4250807}%
  \BibitemOpen
  \bibfield  {author} {\bibinfo {author} {\bibfnamefont {O.}~\bibnamefont
  {Fornieri}},\ }\href {https://doi.org/10.5281/zenodo.4250807} {\enquote
  {\bibinfo {title} {{ottaviofornieri/Diffusion\_MHD\_modes: Diffusion
  coefficient from MHD modes}},}\ } (\bibinfo {year} {2020})\BibitemShut
  {NoStop}%
\bibitem [{\citenamefont {{Draine}}(1978)}]{1978ApJS...36..595D}%
  \BibitemOpen
  \bibfield  {author} {\bibinfo {author} {\bibfnamefont {B.~T.}\ \bibnamefont
  {{Draine}}},\ }\href {\doibase 10.1086/190513} {\bibfield  {journal}
  {\bibinfo  {journal} {\apjs}\ }\textbf {\bibinfo {volume} {36}},\ \bibinfo
  {pages} {595} (\bibinfo {year} {1978})}\BibitemShut {NoStop}%
\bibitem [{\citenamefont {{Wesson}}\ \emph {et~al.}(2005)\citenamefont
  {{Wesson}}, \citenamefont {{Liu}},\ and\ \citenamefont
  {{Barlow}}}]{2005MNRAS.362..424W}%
  \BibitemOpen
  \bibfield  {author} {\bibinfo {author} {\bibfnamefont {R.}~\bibnamefont
  {{Wesson}}}, \bibinfo {author} {\bibfnamefont {X.~W.}\ \bibnamefont {{Liu}}},
  \ and\ \bibinfo {author} {\bibfnamefont {M.~J.}\ \bibnamefont {{Barlow}}},\
  }\href {\doibase 10.1111/j.1365-2966.2005.09325.x} {\bibfield  {journal}
  {\bibinfo  {journal} {\mnras}\ }\textbf {\bibinfo {volume} {362}},\ \bibinfo
  {pages} {424} (\bibinfo {year} {2005})}\BibitemShut {NoStop}%
\bibitem [{\citenamefont {{Heiles}}\ and\ \citenamefont
  {{Crutcher}}(2005)}]{2005LNP...664..137H}%
  \BibitemOpen
  \bibfield  {author} {\bibinfo {author} {\bibfnamefont {C.}~\bibnamefont
  {{Heiles}}}\ and\ \bibinfo {author} {\bibfnamefont {R.}~\bibnamefont
  {{Crutcher}}},\ }\enquote {\bibinfo {title} {{Magnetic Fields in Diffuse HI
  and Molecular Clouds}},}\ in\ \href {\doibase 10.1007/11369875\_7} {\emph
  {\bibinfo {booktitle} {Cosmic Magnetic Fields}}},\ Vol.\ \bibinfo {volume}
  {664},\ \bibinfo {editor} {edited by\ \bibinfo {editor} {\bibfnamefont
  {R.}~\bibnamefont {{Wielebinski}}}\ and\ \bibinfo {editor} {\bibfnamefont
  {R.}~\bibnamefont {{Beck}}}}\ (\bibinfo {year} {2005})\ p.\ \bibinfo {pages}
  {137}\BibitemShut {NoStop}%
\bibitem [{\citenamefont {Langtangen}\ and\ \citenamefont
  {Linge}(2017)}]{langtangen2017finite}%
  \BibitemOpen
  \bibfield  {author} {\bibinfo {author} {\bibfnamefont {H.~P.}\ \bibnamefont
  {Langtangen}}\ and\ \bibinfo {author} {\bibfnamefont {S.}~\bibnamefont
  {Linge}},\ }\href@noop {} {\emph {\bibinfo {title} {Finite difference
  computing with PDEs: a modern software approach}}}\ (\bibinfo  {publisher}
  {Springer Nature},\ \bibinfo {year} {2017})\BibitemShut {NoStop}%
\bibitem [{\citenamefont {Sampson}\ \emph {et~al.}(2022)\citenamefont
  {Sampson}, \citenamefont {Beattie}, \citenamefont {Krumholz}, \citenamefont
  {Crocker}, \citenamefont {Federrath},\ and\ \citenamefont
  {Seta}}]{Sampson:2022cju}%
  \BibitemOpen
  \bibfield  {author} {\bibinfo {author} {\bibfnamefont {M.~L.}\ \bibnamefont
  {Sampson}}, \bibinfo {author} {\bibfnamefont {J.~R.}\ \bibnamefont
  {Beattie}}, \bibinfo {author} {\bibfnamefont {M.~R.}\ \bibnamefont
  {Krumholz}}, \bibinfo {author} {\bibfnamefont {R.~M.}\ \bibnamefont
  {Crocker}}, \bibinfo {author} {\bibfnamefont {C.}~\bibnamefont {Federrath}},
  \ and\ \bibinfo {author} {\bibfnamefont {A.}~\bibnamefont {Seta}},\
  }\href@noop {} {\  (\bibinfo {year} {2022})},\ \Eprint
  {http://arxiv.org/abs/2205.08174} {arXiv:2205.08174 [astro-ph.GA]}
  \BibitemShut {NoStop}%
\bibitem [{\citenamefont {Beattie}\ \emph {et~al.}(2022)\citenamefont
  {Beattie}, \citenamefont {Krumholz}, \citenamefont {Federrath}, \citenamefont
  {Sampson},\ and\ \citenamefont {Crocker}}]{10.3389/fspas.2022.900900}%
  \BibitemOpen
  \bibfield  {author} {\bibinfo {author} {\bibfnamefont {J.~R.}\ \bibnamefont
  {Beattie}}, \bibinfo {author} {\bibfnamefont {M.~R.}\ \bibnamefont
  {Krumholz}}, \bibinfo {author} {\bibfnamefont {C.}~\bibnamefont {Federrath}},
  \bibinfo {author} {\bibfnamefont {M.~L.}\ \bibnamefont {Sampson}}, \ and\
  \bibinfo {author} {\bibfnamefont {R.~M.}\ \bibnamefont {Crocker}},\ }\href
  {\doibase 10.3389/fspas.2022.900900} {\bibfield  {journal} {\bibinfo
  {journal} {Frontiers in Astronomy and Space Sciences}\ }\textbf {\bibinfo
  {volume} {9}} (\bibinfo {year} {2022}),\
  10.3389/fspas.2022.900900}\BibitemShut {NoStop}%
\bibitem [{\citenamefont {{Dame}}\ \emph {et~al.}(2001)\citenamefont {{Dame}},
  \citenamefont {{Hartmann}},\ and\ \citenamefont {{Thaddeus}}}]{COcompare}%
  \BibitemOpen
  \bibfield  {author} {\bibinfo {author} {\bibfnamefont {T.~M.}\ \bibnamefont
  {{Dame}}}, \bibinfo {author} {\bibfnamefont {D.}~\bibnamefont {{Hartmann}}},
  \ and\ \bibinfo {author} {\bibfnamefont {P.}~\bibnamefont {{Thaddeus}}},\
  }\href {\doibase 10.1086/318388} {\bibfield  {journal} {\bibinfo  {journal}
  {\apj}\ }\textbf {\bibinfo {volume} {547}},\ \bibinfo {pages} {792} (\bibinfo
  {year} {2001})},\ \Eprint {http://arxiv.org/abs/astro-ph/0009217}
  {arXiv:astro-ph/0009217 [astro-ph]} \BibitemShut {NoStop}%
\end{thebibliography}%
\bibliographystyle{apsrev4-1}
\end{document}